\documentclass[conference]{IEEEtran}

\usepackage{cite}      

\usepackage[usenames,dvipsnames,svgnames,table]{xcolor}
\usepackage{graphicx}  

\usepackage{psfrag}    

\usepackage{url}       
\usepackage{tikz}

\usepackage{amsmath}   
\usepackage{amssymb}
\usepackage{algorithm}
\usepackage{algpseudocode}
\usepackage{pifont}

\usepackage{tikz}
\usetikzlibrary{shapes,arrows}
\usepackage[T1]{fontenc}
\usepackage{listings}
\usepackage{subfigure} 
\usepackage{scrextend}

\newcommand{\shorten}[1]{}

\newtheorem{lemma}{Lemma}

\newcommand{\signed}%
    {{\unskip\nobreak\hfill\penalty50
      \hskip2em\hbox{}\nobreak\hfil $\blacksquare$
      \parfillskip=0pt \finalhyphendemerits=0 \par}}

\IEEEoverridecommandlockouts

\begin{document}

\title{Local Voting: Optimal Distributed Node Scheduling Algorithm for Multihop Wireless Networks}

%
\author{
	\IEEEauthorblockN{Dimitrios~J.~Vergados\IEEEauthorrefmark{1}, Natalia~Amelina\IEEEauthorrefmark{2}, Yuming~Jiang\IEEEauthorrefmark{3}, Katina Kralevska\IEEEauthorrefmark{3}, and~Oleg~Granichin\IEEEauthorrefmark{2}\\ Email: djvergad@gmail.com, n.amelina@spbu.ru, \{jiang, katinak\}@item.ntnu.no, o.granichin@spbu.ru}
	\IEEEauthorblockA{\IEEEauthorrefmark{1}School of Electrical and Computer Engineering, National Technical University of Athens}
	\IEEEauthorblockA{\IEEEauthorrefmark{2}Faculty of Mathematics and Mechanics, Saint-Petersburg State University}
	\IEEEauthorblockA{\IEEEauthorrefmark{3}Department of Information Security and Communication Technology, Norwegian University of Science and Technology}
	\thanks{The work of N. Amelina and O. Granichin was supported by RFBR (Grants No.15-08-02640 and No.16-01-00759).}
}



\maketitle

\begin{abstract}
	An efficient and fair node scheduling is a big challenge in multihop wireless networks. In this work, we propose a distributed node scheduling algorithm, called \emph{Local Voting}.
	The idea comes from the finding that the shortest delivery time or delay is obtained when the load is equalized throughout the network.
	Simulation results demonstrate that \emph{Local Voting} achieves better performance in
	terms of average delay, maximum delay, and fairness compared to several representative scheduling
	algorithms from the literature. Despite being distributed, \emph{Local Voting} has a very close performance to a centralized algorithm that is considered to have the optimal performance.


\end{abstract}


%
\IEEEpeerreviewmaketitle

\section{Introduction} \label{intro}
Node scheduling algorithms in wireless networks assign each transmission opportunity to a set of nodes in a way that ensures that there is no mutual interference among any transmitting nodes.
Some representative algorithms from the literature are DRAND~\cite{rhee2006drand}, Lyui's algorithm~\cite{hammond2004properties}, Load-Based Transmission Scheduling
(LoBaTS)~\cite{wolf2006distributed}, and Longest Queue First
(LQF)~\cite{dimakis2006sufficient}. These algorithms are distributed except LQF, and they have different specifications in terms of traffic and topology dependence.

We propose a distributed, traffic and topology dependent algorithm called \emph{Local Voting}.
This algorithm tries to equalize the load (defined as the ratio of the queue
length and the number of allocated slots) through slot reallocation based on
local information exchange between neighboring nodes.
The number of reallocated slots is determined by the relation between the load of each node
and its neighbors, under the limitation that certain slot exchanges are not
possible due to interference with other nodes.
\emph{Local Voting} enables nodes with lower load to give slots to nodes with higher load, and, eventually the load between nodes reaches a common value, i.e. consensus is achieved \cite{7047923}.

\section{Network model and Optimal Strategy} \label{model}
Consider a network of $n$ nodes where $N = \{1,2,\ldots,n\}$ is the set of all wireless nodes that communicate over a shared wireless channel.
 The channel access follows a paradigm of time division multiple access. Time is divided into frames $t=1, 2, \dots$, and each frame is divided into $S$ slots where the duration of a time slot is sufficient to transmit a single packet. There is no spatial movement of the nodes.

The transmission schedule of the network is defined as,
\begin{equation}
\label{Djv_01}
X_t^{i,s} = \left\{ \begin{array}{rl}
1, &\mbox{ if a slot } s \in S \mbox{ is assigned to a node } i \in N\\
0, &\mbox{ otherwise}
\end{array} \right.
\end{equation}
for $t \ge 1$, with $X^{i,s}_0 = 0$  by convention.

The transmission schedule is \emph{conflict-free}, if for any $t$,
\begin{equation}
\label{Djv_1}
X_t^{i,s} X_t^{j,s} = 0, \forall s \in S, i \in N, j \in N_{i}^{(2)}, i \neq j
\end{equation}
where $N_{i}^{(2)}$ denotes the \emph{two-hop neighborhood} of node $i$, i.e. {\em the set
	of all nodes that are neighbors to node $i$ or that have a common neighbor with node $i$}. If we define $N_{i}^{(1)}$ as \emph{one-hop neighborhood} of node $i$, there holds $N_{i}^{(1)} \subset N_{i}^{(2)}$.

The objective is to design a load-balancing node scheduling strategy that schedules transmissions such that minimum maximal (minmax) delivery time or delay is achieved.

At any time $t$, the state of each node $i$ is described by:
\begin{itemize}
	\item $q_{t}^{i} $ is the queue length, counted as the number of slots needed
	to transmit all packets at node $i$ at time $t$;
	\item $p_{t}^{i} $ is the number of slots assigned to node $i$ at time $t$,
	i.e. $p_{t}^{i} = \sum\limits_{s=1}^{|S|} X_t^{i,s} $.
\end{itemize}

The dynamics of each node is described by
\begin{eqnarray}
\label{Nat_11}
\begin{aligned}
q_{t + 1}^{i} &= \max\{0, q_{t}^{i} - p_{t}^{i}\} + z_{t+1}^{i}, \; i \in N,
\; t = 0,  1, \ldots, \\
p_{t + 1}^{i} &= p_{t}^{i} + u_{t+1}^{i},
\end{aligned}
\end{eqnarray}
where $z_{t}^{i}$ is the number of slots needed to transmit new packets received by node $i$ at time $t$, and $u_{t+1}^{i}$ is the number of slots that node $i$ gains or loses at $t+1$ due to the adopted strategy.

\begin{lemma}[Optimal strategy]\label{lemma_optim}
	Among all possible options for load balancing, minmax completion time is
	achieved when
	\begin{equation}
	p_t^i/\max\{1, q_t^i\} = p_t^j/\max\{1, q_t^j\}, \; \forall i, j \;\in N.
	\end{equation}
\end{lemma}

\section{Local Voting Algorithm} \label{LV}
Corresponding to the optimal strategy, if we take $x_t^i=p_t^i/\max\{1, q_t^i\}$ as the state
of node $i$, then the goal is to keep this ratio
equal, i.e. load balancing or $x_t^i = x_t^j$ for all $i, j \in N$.

\begin{figure}
	\centering
	\includegraphics[width=3in]{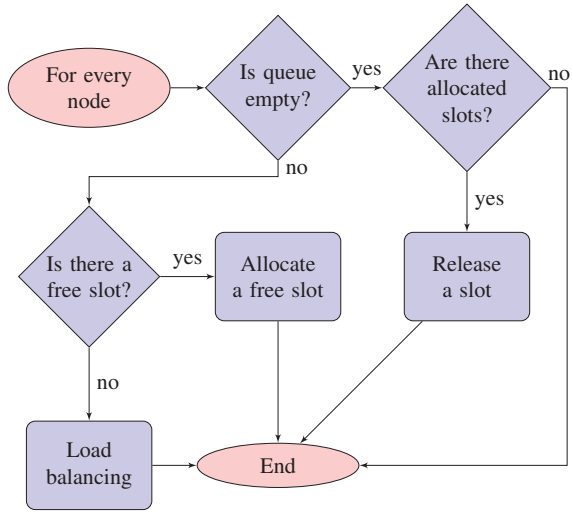}
	\vspace{-0.3cm}
	\caption{Requesting and releasing time slots function}
	\label{fig:algo1}
\end{figure}

The proposed \emph{Local Voting} algorithm consists of two main functions:
requesting and releasing time slots, and load balancing.
For the first function (Fig.~\ref{fig:algo1}), nodes are examined sequentially at the beginning of each frame. If the queue of a node is not empty, then the first available slot which is not reserved by one-hop or two-hop neighbors is allocated to the node.
If no available slot is found, then no slot is allocated to the node. On the contrary, if the queue is empty and the node has allocated slots,
then one of the slots is released.

\begin{figure}
	\centering
	\includegraphics[width=2.3in]{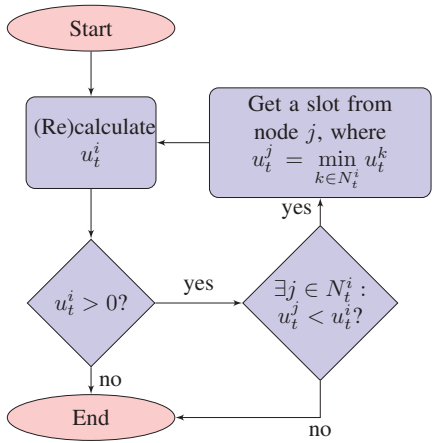}
	\vspace{-0.3cm}
	\caption{Load balancing function}
	\label{fig:algo2}
	\vspace{-0.2cm}
\end{figure}

The load balancing function (Fig.~\ref{fig:algo2}) is invoked whenever a
node has a non-empty queue and no free slots are available.
Each node calculates a value $u_t^i$ (as explained in~\cite{DBLP:journals/corr/VergadosAJKG17}), which determines how many slots the node should ideally gain or release by the load balancing function.
If a node has a
positive $u_t^i$ value, then it checks if any of its neighbors, which may give a slot to it without causing a conflict, has a $u_t^j$ value smaller than the $u_t^i$ value of the requesting node $i$.
The neighbor with the smallest $u_t^j$ value gives a slot
to node $i$. After the exchange, $u_t^j$ is reduced by one, and $u_t^i$ is recalculated. This procedure is repeated until $u_t^i$ is not positive, or
until none of the neighbors of node $i$ can give slots to node $i$ without causing a
conflict.

\begin{figure}
	\centering
	\includegraphics[width=3in]{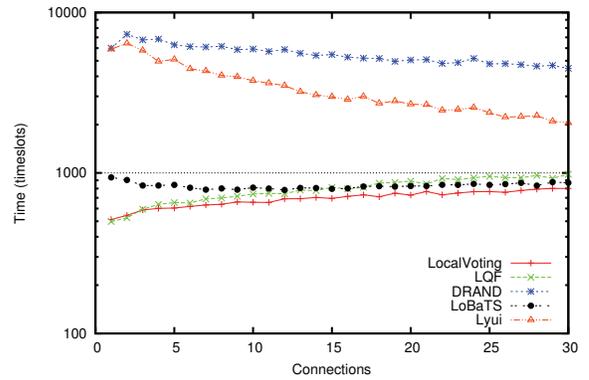}
	\vspace{-0.3cm}
	\caption{The minimum end-to-end delay}
	\label{fig:min}
\end{figure}

\begin{figure}
	\centering
	\includegraphics[width=3in]{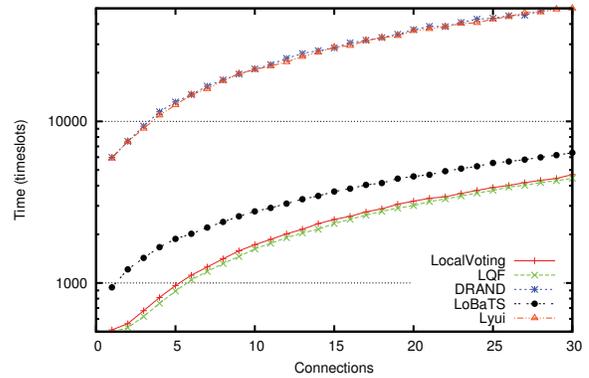}
	\vspace{-0.3cm}
	\caption{The maximum end-to-end delay}
	\vspace{-0.2cm}
	\label{fig:max}
\end{figure}
\section{Results and Future Work} \label{results}
Fig. \ref{fig:min} and Fig. \ref{fig:max} demonstrate the minimum and maximum end-to-end delays, respectively. For each connection, 100 equally sized packets are generated at regular intervals (every $5$ slots) at the source and forwarded towards
the destination. The source and the destination are chosen
randomly. The simulation ends after all traffic has reached its destination. The delay is minimized with \emph{Local Voting} compared to DRAND~\cite{rhee2006drand}, Lyui ~\cite{hammond2004properties}, LoBaTS~\cite{wolf2006distributed}, and LQF~\cite{dimakis2006sufficient}.
Although, \emph{Local Voting} is a distributed algorithm, the maximum end-to-end delay is very close to the centralized \emph{LQF} algorithm.
We plan to apply \emph{Local Voting} in Internet of Things and resource balancing in 5G networks.
\vspace{-0.2cm}
\bibliographystyle{IEEEtran}
\bibliography{SchedulingConsensus}

\end{document}